\documentstyle[graphicx,preprint,prb,aps]{revtex}
\begin{document}
\draft
\title{Multipole modes and spin features in the Raman
spectrum of nanoscopic quantum rings}
\author{Agust\'{\i} Emperador, Mart\'{\i} Pi, and Manuel Barranco}
\address{Departament ECM, Facultat de F\'{\i}sica,
Universitat de Barcelona, E-08028 Barcelona, Spain}
\author{Enrico Lipparini}
\address{Dipartimento di Fisica, Universit\`a di Trento,
and INFM sezione di Trento, I-38050 Povo, Italy}
\date{\today}

\maketitle

\begin{abstract}

We present a systematic study of ground state and spectroscopic
properties of many-electron nanoscopic quantum rings.
Addition energies at zero magnetic field $(B)$ and electro-chemical
potentials  as a function of $B$ are given for a ring hosting
up to 24 electrons. We find discontinuities in the excitation energies
of multipole spin and charge density modes, and a coupling between
the charge and spin density responses
that allow to identify the formation of ferromagnetic ground
states in narrow magnetic field regions. These effects
can be observed in Raman experiments, and are related
to the fractional Aharonov-Bohm oscillations of 
the energy and of the persistent current in the ring.

\end{abstract}
\narrowtext
\section{Introduction}

Very recently, quantum rings in InAs-GaAs heterostructures have been
fabricated in the nanometer scale \cite{Gar97,Lor98,Lor00},
and the capacitance and far-infrared (FIR) response have been
measured for the one- and two-electron quantum rings. Previously,
the FIR response\cite{Dah93} had been measured for mesoscopic rings
in GaAs-Ga$_x$Al$_{1-x}$As heterostructures, for which a description
based on classical and hydrodynamical models works fairly
well\cite{Pro92,Zar96}.
In nanoscopic rings quantum effects are important and for this reason
theoretical studies  at a more microscopic level have been undertaken
\cite{Cha94,Gud94,Wen96,Hal96,Nie96,Tan99,Vie00,Mag99,Emp99,Hu00,Kos00,Bor00,Pue01}.

In some of these calculations a major emphasis has been put on
describing the Aharonov-Bohm (AB) quantum effect which manifests
in the presence of an external magnetic field $(B)$
as the existence of a persistent current, and leads to periodic
oscillations in the energy spectrum and the persistent current as a
function of $B$. These AB oscillations  have been observed in
mesoscopic rings in a GaAlAs/GaAs heterostructure by
measuring the conductance across the ring\cite{Mai93}.

The experimental results for one- and two-electron nanoscopic rings have
been theoretically analyzed\cite{Emp00} using the 
current density (CDFT) and time-dependent
local-spin density-functional (TDLSDFT) theories, and a good
agreement between theory and experiment is found (see also
Refs. \onlinecite{Hu00} and
\onlinecite{Pue01}). Motivated by this success,
in this work we extend this approach  to a systematic study of
ground state (gs)
and spectroscopic properties of nanoscopic rings containing up to
$N=24$ electrons. Our aim is to show the physical appearance of
quantities that could be measured in the future if
the problem of fabricating many-electron nanoscopic rings is
eventually solved. Specifically, we have obtained
electro-chemical potentials $\mu(N)=E(N)-E(N-1)$,
where $E(N)$ is the total energy of the $N$-electron ring,
addition energies
$\Delta_2(N)=\mu(N+1)-\mu(N) = E(N+1) - 2 E(N) + E(N-1)$,
and the spin and charge density responses at finite on-plane
transferred wave-vector,
which are relevant for the analysis of Raman spectra.
All these quantities have been measured in $N$-electron quantum
dots, see for example Refs.
\onlinecite{McE91,Ash92,Tar96,Loc96,Sch96,Sch98}
and references therein.
Particular emphasis is given to  spectroscopic results.
We show that  the collective excitation  energies exhibit
discontinuities in their $B$ dependence, which are a manifestation of
changes in the spin configuration of the ground state,
and that in a ring-wire would be related to the fractional
Aharonov-Bohm effect.

The structure of the ring gs has been obtained within the CDFT as
described in Refs. \onlinecite{Fer94} and \onlinecite{Pi98}.
To obtain the charge and spin density responses
we have used
TDLSDFT as described in Refs. \onlinecite{Ser99} and
\onlinecite{Bar00}.
Essentially, the method implies that to obtain the
ground state and excited modes of the system, besides the direct
electron-electron interaction, exchange 
and correlation effects have been included in a local density
approximation.  We refer the reader to these references
for a comprehensive exposure of the CDFT and TDLSDFT, of direct
applicability here changing the shape of the confining potential
from a dotlike potential to a ringlike potential.

\section{Ground state results}

Following Ref. \onlinecite{Cha94}, we have modeled the 
ring
confining potential by a parabola
\begin{equation}
V^+(r)=\frac{1}{2}m\,\omega_0^2\,(r-R_0)^2~
\label{eq1}
\end{equation}
with $R_0=20$ nm and  $\omega_0=12$ meV. These values are close to
the ones used\cite{Emp00} to describe the rings studied by
Lorke et al\cite{Lor00}. The electron
effective mass $m^*=0.063$  (we write $m = m^* m_e$ with $m_e$
being the physical electron mass) and effective gyromagnetic
factor $g^*=-0.43$ have
been taken from the experiments\cite{Fri96,Lor96,Mil97}, and
the value of the dielectric constant has been taken to
be $\epsilon=12.4$. The model is strictly two-dimensional, and
as a consequence of circular symmetry, the single particle
(sp) wave functions are
eigenstates of the orbital angular momentum $l_z$ and can be written
as $u_{nl\sigma}(r) e^{-il\theta}$ with $l =0,\pm1,\pm2\ldots$ Notice
that with this convention, the sp orbital angular momentum is $-l$.

Some times we have used effective atomic units,
defined by $\hbar=e^2/\epsilon=m=$1.
In this system of units, the length unit is the effective
Bohr radius $a_0^* = a_0\epsilon/m^*$,
and the energy unit is the effective Hartree
$H^* = H  m^*/\epsilon^2$.
This yields  $a^*_0 \sim$ 10.4 nm and $H^*\sim$ 11.15 meV.
The  Bohr magneton  is defined as $\mu _B=\hbar e/2 m_e c$.

Fig. \ref{fig1} shows the addition energies $\Delta_2(N)$
 at zero magnetic field.
Saw-tooth structures and large peaks at
magic numbers $N=2$, 6, 10, 16 and 24 that correspond to
close-shell rings are clearly seen. Had we not taken into
account electron-electron
interactions (single-particle picture), the magic
numbers would have been $N=2$, 6, 10, 16, 20 and 24.
The effect of interaction is to weaken some shell
closures, for example that at $N=20$ in the displayed results.
The total-spin third-component of the gs configuration is
indicated in the figure,
and shows that Hund's first rule, according to which
as degenerate states are filled, the total spin $S$
takes the maximum allowed value
by the exclusion principle and becomes zero for
closed shells, is satisfied up to $N=16$.
For larger electron numbers Hund's first rule is violated
and the higher spin  state no longer is the gs.
In particular, $S_z$ becomes zero at
$N=20$ instead of having the Hund's first rule value $S_z=2$.
These considerations may
change if the geometrical characteristics
of the studied ring are different. In this respect,
we  want to
point out that according to the one-dimensionality
criterium established in Ref. \onlinecite{Vie00},
the rings we describe cannot be considered as
ring-wires, an obvious statement if one looks at
the electron density plotted in Fig. \ref{fig2}.

Similarly to what happens in quantum dots,\cite{Ste98}
the application of an external  magnetic field
produces a nontrivial change in the ring structure,
inducing changes in the spin of the gs
configuration. This can be seen
in the middle panel of Fig. \ref{fig3}
where we have plotted the $B$ evolution
of the sp energies up to $B=6.5$ T for $N=10$.
Paying attention to the  $4.8 < B < 6$ T region for example,
one may observe a level crossing around the Fermi energy
when $B\approx 5$ T.  An
electron moves from the $(0,\downarrow)$  to
the $(5,\uparrow)$ sp level and, as a consequence,
the spin state changes from a singlet to a triplet
state (ST transition). In such a way the system
minimizes its  exchange energy and the
gs becomes ferromagnetic. A further increase
of $B$ changes again the situation, and
at $B\approx 5.8$T an electron moves now from the $(0,\uparrow)$
to the $(5,\downarrow)$ sp level and the system undergoes
a triplet $\rightarrow$ singlet (TS) transition.
Exchange gaps in the sp energies similar to this one
also occur near $B \sim $ 1 and 3 T, and again
at  $B\approx 8.5$ T (not shown in the figure),
and are characteristic of even-$N$ rings before
full magnetization. Exchange gaps
have been analyzed in quantum dots to experimentally
determine the contribution form the direct Coulomb
and exchange electron-electron interaction to the
total energy\cite{Tar00}.

The situation is quite
different for odd-$N$ rings, for which the system
always remains ferromagnetic ($S_z\ne0$).
This can be seen in the top panel of Fig. \ref{fig4},
where we have plotted the $B$ evolution of the sp levels
of the $N=$ 9 ring up to $B=6.5$ T. Now,
the level crossings around the Fermi energy
occur without spin changing.
The situation shown in this figure
is similar for all the odd-$N$ rings considered in this work.

 Level crossings strongly
influence the $B$ behavior  of the electro-chemical
potential, which is shown in Fig. \ref{fig5}
for $0 \leq B \leq  10$ T up to $N=12$. For even
electron numbers the number of ST transitions
depends on $N$ and increases with it, as the number of
level crossings to arrive at  full polarization also increases.
The critical $B$ value  at which the system is fully
polarized also increases with $N$,  and is well visible in
this figure up to $N =9$. For odd electron numbers,
the spin of the gs configuration remains constant
($S_z=\slantfrac{1}{2}$) in a  wide $B$-range, and from  a certain
$B$ value on it starts increasing up to
full polarization.

The pairing between peaks in the electrochemical potential
corresponding to consecutive $N$ values, as for example
9 and 10, indicates antiparallel spin filling
of a sp $l$-orbital by two electrons. This pairing is
violated at $B=0$  for $N=4$ and 8, and it is a manifestation
of Hund's first rule. It is also violated in the narrow
$B$-regions where the spin triplet state is the gs configuration
of even-$N$ rings. This effect
cannot be seen in Fig. \ref{fig5} because of the
$\Delta B= 0.5$ T step used to obtain it, but can be seen in
the top panel of Fig. \ref{fig11} below, which gives a
magnified view of $\mu(10)$ in the region around $B\sim 3$ T.
This is the same behavior found for quantum dots
in Ref. \onlinecite{Tar00}.

Quasi-periodic $B$-oscillations can be observed in the
electro-chemical potential.
The changes in spin configurations and the maxima and
minima of the electro-chemical potential curves
are related to sp level crossing at
the Fermi energy. For example,  for $N=10$
the ST transitions around $B=1$, 3, and 5.5 T,
 and the maxima of $\mu(10)$,
can be related to the level crossings at these $B$
values shown in Fig. \ref{fig3}. The minima of $\mu(9)$
and $\mu(10)$ at $B\sim2$ and 4.5 T correspond
to the level crossings without spin change shown in
Figs. \ref{fig3} and \ref{fig4}.

These quasi-periodic oscillations
are absent in quantum dots,\cite{Tar96}
and are a characteristic inherent to quantum rings.
We recall that for a one-electron ring-wire
 of radius $R_0$ the energies are given by
\begin{equation}
\epsilon_l=\frac{\hbar^2}{2m^*
R_0^2}\left(l-\frac{\Phi}{\Phi_0}\right)^2 ~,
\label{eq2}
\end{equation}
where $\Phi/\Phi_0$ is the number of flux
quanta penetrating the ring,
$\Phi_0=h c/e$, and are periodic in the flux
$\Phi=\pi R_0^2 B$
with periodicity $\Phi_0$. The persistent current
$I_l$ associated with the $l$-state
\begin{equation}
I_l=-c\frac{\partial \epsilon_l}{\partial \Phi}
\label{eq3}
\end{equation}
has the same periodicity. As $B$ increases the occupied
$l$-state changes and  gives rise to discontinuities
in the current $I_l$ (the AB effect), which have
been experimentally observed in a mesoscopic loop in a GaAs
heterojunction\cite{Mai93}.

We have plotted in Figs. \ref{fig3} and  \ref{fig4}
the total energy
$E$ and
the persistent current determined as $I=\int J(r)\,dr$
with
\begin{equation}
{\bf J(r)}= {\bf J}_p(r) - \frac{e}{m^*c}\,n(r){\bf A}(r)\; ,
\label{eq4}
\end{equation}
where 
${\bf J}_p(r)= -\hat{e}_{\theta}\sum l[u_{nl\sigma}(r)]^2/r$,
being $\hat{e}_{\theta}$ the azimuthal vector unit,
is the paramagnetic current that for many-electron
rings is obtained adding the sp paramagnetic currents,\cite{Vig87}
 and the second term is the diamagnetic
current, being $n(r) = \sum [u_{nl\sigma}(r)]^2$
and ${\bf A}(r)$ the electron density
and external vector potential, respectively.

In a many-electron nanoscopic ring-wire, electron-electron
interactions give rise to changes in the gs configuration
that decreases the period of the AB oscillations\cite{Mos00}.
This effect has been explained in Ref. \onlinecite{Nie96}
as a consequence of Hund's first rule which favors the occurence
of ferromagnetic $S_z=1$ phases.
A ring with a low degree of one-dimensionality tends to loose
the periodic features of the ideal ring-wire, being somehow
an intermediate case between a ring-wire and a quantum dot,
and for these systems the effect is not so marked,\cite{Emp00}
appearing as a small amplitude oscillation whose $B$-width
coincides with that of the ferromagnetic phase
(see Fig. \ref{fig3}).
The crucial role played by the degree of one-dimensionality in the
appearance of the AB oscillations can also be appraised
comparing the results for two and four electron rings obtained in
Refs. \onlinecite{Nie96,Vie00}.
In the case of an $S_z=0$ ring as for $N=10$, we have found
a period halving.
For an $S_z=\slantfrac{1}{2}$ ring like that with $N=9$, the 
oscillations
due to ST and TS transitions are absent but there still is
a period halving which in this case can be traced back to
that arising for  non-interacting
spin-$\slantfrac{1}{2}$ fermions.\cite{Nie96,Mos00} We have estimated
that to have $\Phi = \Phi_0$ for a ring-wire of radius
equal to the maximum electronic density radius in Fig. \ref{fig2},
one needs $B \sim 2.2$ T. Consequently, a period of $\Phi_0/2$
corresponds to a period of $B \sim 1$ T as shown in the bottom
panel of Fig. \ref{fig4}. A period $\sim \Phi_0$, i.e.,
$B \sim 2$ T between large amplitude oscillations of the current
shown in the bottom panel of
Fig. \ref{fig3} is also consisten with this picture.

We have also checked that calculating
the current from the $B$ derivative of the total energy,
 as suggested by Eq. (\ref{eq3}), one obtains discontinuities at
the same $B$ values. However, both ways of calculating the persistent
current do not yield the same value. This difference is attributed
to the fact that our ring is not in a `clean' AB situation in which
electrons move essentially in a $B$-free region\cite{Vie00}. Rather,
an appreciable magnetic flux is going across the effective ring surface.

\section{Raman spectra}

Raman spectra are obtained through two-photon inelastic
light scattering.
The process consists in creating an electron-hole pair
(absorption of the incident photon) and subsequent 
recombination between valence and conduction bands
(emission of the final photon).
In this process\cite{Pin89,Stei99}
the electrons of the conduction band are excited via
the transfer of energy $\omega$ and momentum $q$.
In a backscattering geometry, with incoming and scattered
photon polarization vectors $\hat{\bf e}_i$ and
$\hat{\bf e}_s$ lying in the $x-y$ plane perpendicular to $B$,
the charge  and spin density  Raman cross sections
are essentially determined by the longitudinal strength
(or dynamical structure) functions\cite{Bar00,Blu70,Ste99}
in the charge (CDE)  $S_{nn}(q,\omega)$ and the spin
density channel (SDE) $S_{mm}(q,\omega)$:
\begin{eqnarray}
\frac{d^2\,\sigma^{C}}{d\omega_s\, d\Omega_s} &\propto& |\hat{\bf e}_i
\cdot \hat{\bf e}_s|^2 S_{nn}(q,\omega)
\nonumber
\\
& &
\label{eq5}
\\
\frac{d^2\,\sigma^{S}}{d\omega_s\, d\Omega_s} &\propto& |\hat{\bf e}_i
\times \hat{\bf e}_s|^2 S_{mm}(q,\omega) \,\,\, ,
\nonumber
\end{eqnarray}
where  $\omega = \omega_i - \omega_s$ is the energy difference
of the incoming and scattered photon. Using the above expressions
one assumes that only conduction band electron are involved, and
only off-resonance Raman peaks excited
by laser energies above the valence-conduction
band gap can be described.

To obtain $S_{nn}(q,\omega)$ and  $S_{mm}(q,\omega)$ within the
TDLSDFT, we have calculated  the response to operators whose
spatial dependence in the on-plane wave vector $\vec{q}$
is a plane wave $e^{i \vec{q}\,\vec{r}}$.
We refer the reader to Refs. \onlinecite{Ser99} and
\onlinecite{Bar00} for a detailed
description of the longitudinal response at $q \approx 0$, and
at finite $q$.

Figure \ref{fig6} shows the CDE's and SDE's at $B=0$ for the $N=10$ ring
and several $q$ values. For each $q$ value, we have considered in the
expansion of the plane wave into Bessel functions of integer order
\begin{equation}
e^{i \vec{q}\,\vec{r}} =
 J_0(qr) +
\sum_{L>0} i^L \,J_L(q r)\, (e^{i L \theta}+
 e^{-i L \theta})
\label{eq7}
\end{equation}
as many multipole terms as needed to exhaust the $f$-sum rule\cite{Bar00}:
\begin{equation}
m_1^{(nn)}[e^{i \vec{q}\, \vec{r}}\,] =
m_1^{(mm)}[e^{i \vec{q}\, \vec{r}}\,] = q^2 \frac{N}{2}\; .
\label{eq8}
\end{equation}
We infer from this figure that the modes have no appreciable wave-vector
dispersion, a clear signature of the `zero-dimensionality' of the ring.
A similar conclusion was drawn for dots,\cite{Bar00} in
agreement with experiment.\cite{Sch96} It can also be observed from this
figure that a few number of multipoles (up to $L=2$ for the
largest $q$ value) is enough to yield the plane wave response, and that
in the $q \approx 0$ long wave-length limit only the dipole
$L=1$ mode contributes.

The $B$ dispersion of the dipole mode for
the $N=10$ ring in the long wave-length limit
is shown in the top panel of Fig. \ref{fig3}.
It is worth to see that edge modes have
a smooth $B$ dispersion except in the regions where the ring is in the
ferromagnetic $S_z=1$ phase. The discontinuities in the magnetoplasmon
and spin modes, or equivalently
these appearing in the absorption spectrum,\cite{Nie96} are
features that could be detected in optical-absorption experiments,
making observable the ST and TS transitions and the ferromagnetic
phases.

In the one-electron ring-wire case, the energies of the
dipole modes are given by

\begin{eqnarray}
\omega_+=\epsilon_{l+1}-\epsilon_l=\frac{\hbar^2}{2m^*R_0^2}
\left[1-2\left(\frac{\Phi}{\Phi_0}-l\right)\right]
\nonumber
\\
& &
\label{eq9}
\\
\omega_-=\epsilon_{l-1}-\epsilon_l=\frac{\hbar^2}{2m^*R_0^2}
\left[1+2\left(\frac{\Phi}{\Phi_0}-l\right)\right]~.
\nonumber
\end{eqnarray}
These excitation energies show AB oscillations
with the same period as gs energy and persistent
current oscillations. Furthermore,
they are discontinuous at the $B$ values corresponding to
level crossings.
The $B$ dependence of the absorption spectrum of a quasi ring-wire
with two electrons was obtained\cite{Nie96}
by an exact diagonalization calculation, and it was shown to
display fractionary AB oscillations of  period $\Phi_0/2$.
The ring-wire character of the system is crucial for it to present
distinct AB oscillations.

Neither spin nor charge density modes with $L \neq 1$
can be detected in FIR spectroscopy.
In contradistinction, Raman spectroscopy in QD's has proved to be
able to disentangle spin from density modes, and to identify
non-dipole charge and spin density modes.\cite{Sch98} We show the charge
and spin density strengths for $q=0.05$ nm$^{-1}$ in Figs.
\ref{fig7} and \ref{fig8}, respectively. Only modes up to the octupole
($L=3$) mode are excited with some intensity.
The octupole strength is very small because at $q=0.05$ nm$^{-1}$
it contributes very little to the Bessel-function expansion of
the plane wave in the region where
the electronic density  is concentrated, see Fig. \ref{fig2}.
The monopole $L=0$ strength in the charge density channel has only one
high-energy peak, whereas it is very fragmented in the spin density channel.

We have identified the polarization of the main peaks with a $+$ ($-$) sign
when the peak arise from $\Delta |L_z|=L (-L)$ changes in
orbital angular momentum. They  come, respectively,  from the $e^{i L\theta}$
and $e^{- i L\theta}$
contributions in the expansion of the plane wave, Eq. (\ref{eq7}).
The lower energy $(+)$ peak corresponds to the outer edge mode,
the high energy ($-$) peak to the bulk mode, and the lower ($-$) peak
to the inner edge mode.\cite{Emp99,Emp00} Actually, the distinction
between edge and bulk modes only makes sense at high enough magnetic fields;
in the present case, well developed Landau bands appear
at $B \sim 5$ T, and for lower magnetic fields the charge and spin
density spectrum is fairly complex.

We have collected in Figs. \ref{fig9} and \ref{fig10} the $B$
dispersion of the more intense CDE's and SDE's, respectively,
for $q=0.05$ nm$^{-1}$.
It can be noted that the discontinuities in the energy of the inner
edge dipole mode ($-$) we have discussed in the long wave-length limit
(top panel of Fig. \ref{fig3}) are  well visible.
For quadrupole modes, at some $B$ values we have found more than one
single intense peak of a give polarization (see also Fig. \ref{fig7}).
In these cases, the more intense peak is indicated by a full symbol, and
the less intense peak by an open symbol.

At high magnetic fields, the positive $B$ dispersion CDE's and SDE's
have a clear tendency to bundle.
These peaks arise from interband transitions whose Landau index differs in
one unit. As explained in Ref. \onlinecite{Bar00} for
the case of dots, this happens because the Landau bands
are made of many sp states with different $l$ values and energies rather $l$
independent if $B$ is large enough. Only finite-size effects and the $L$
dependence of the electron-hole interaction cause some dependence on the
mode multipolarity. The origin of the high energy, weak quadrupole CDE's
are excitations involving interband transitions whose Landau index
differs in two units. Notice also that the quadrupole spin density
strength is very fragmented, especially around the $ B\sim 5$ T
ferromagnetic region.

It is interesting to notice that `$-$' edge modes only appear in the
dipole case. Recalling that these are inner  edge modes,
Fig. \ref{fig2} helps to understand this finding. It can be seen
that only the $L=0$ and 1 multipoles of the plane wave expansion are
sizeable
in the inner edge region, and as a consequence, the strength of the
$J_L(q r)\, e^{-i L \theta}$ component with $L>1$
goes to the high energy mode of the same polarization. In contradistinction,
the `$+$' outer edge mode has no high-energy  counterpart and takes the whole
strength of the $J_L(q r)\, e^{i L \theta}$ component.
The high sensitivity of the `$-$' edge modes to the ring morphology
might help to shed light into the actual structure of the ring inner
edge\cite{Pue01}. We recall that the experimental nanoscopic 
rings\cite{Lor00} have been fabricated from InAs self-assembled dots
grown on GaAs covered by a thin layer of GaAs by an annealing process.

The existence of ferromagnetic phases causes the coupling of spin and
charge density responses. This effect can be detected
experimentally, and as a matter of fact, it has been observed in QD's
using Raman spectroscopy.\cite{Sch98} This  means
that in the spin density channel, besides the main SDE's,
a spin-dependent external field excites
other low intensity peaks lying at the energy of the
more intense CDE's which appear in the charge density channel, and
conversely. The coupling of CDE's and SDE's
has been thoroughly discussed within TDLSDTF in Refs.
\onlinecite{Ser99} and \onlinecite{Ste99} for QD's, and in
Ref. \onlinecite{Emp01} for antidots. An example of how this coupling
looks like for rings is presented in Fig. \ref{fig11}, where we show
the charge and spin dipole modes near $B \sim 3$ T. One can see that
both responses are decoupled if the gs is paramagnetic, as it is at
$B=2.6$ T, and are coupled if the gs is ferromagnetic, as it is at
$B=2.8$ T, for example.

The existence of islands of spin-triplet
gs's of rather wide $B$-extension is a characteristic of nanoscopic
rings which mesoscopic rings hosting several thousand electrons lack of.
Indeed, only for few electron rings
the spin polarization $\xi = (N_{\uparrow}-N_{\downarrow})/N$ is
sizeable in the triplet state, a $20\%$ for the $N=10$ ring.
Large couplings only arise from large spin polarizations through
the spin dependence of the exchange-correlation potential\cite{Ser99}.

Finally, we present in Fig. \ref{fig12} a grey-scale plot of the
Raman strength in the charge (left panel) and spin density channels
(right panel) corresponding to the $ 4.8 \leq B \leq 6$ T region. This
figure clearly reveals the ST and TS transition points as
discontinuities in the $B$ dispersion of the multipole modes, especially
visible in the inner edge dipole mode. These discontinuities are
fingerprints of gs spin transitions and could be detected in Raman
experiments.

\section{Summary}

In this work we have studied physical aspects of  charge and
spin density responses of nanoscopic rings that might be detected by
Raman spectroscopy. We have considered rings whose morphology is close
to that of the systems recently fabricated, excluding for this reason
the interesting case of nanoscopic ring-wires that can also be addressed
using the same method. We have taken as case of study a ring with
10 electrons and have seen that monopole to quadrupole 
modes clearly appear in the spin, and especially in the charge
density channel. Both channels are coupled if the gs is
ferromagnetic, and the ST and TS transition points may be identified as
discontinuities in the $B$ dispersion of the multipole modes.

We have also discussed the addition spectrum of rings hosting up to 24
electrons, and have found that their shell structure is
similar to that of quantum dots up to $N=6$. For larger $N$ values this
is not so; the rings present shell closures at $N=2, 6, 10, 16$, and 24.
Hund's first rule is satisfied up the fourth shell for the
assumed ring geometry.

The non one-dimensional character of these rings hinders the existence
of fractionary AB oscillations in the sp energy levels, permanent
current and collective modes. Yet, the existence of triplet gs's is
manifested by the correlated presence of spin gaps in the $B$ dipersion
of the sp levels,
discontinuities in the multipole collective modes, and  small
amplitude oscillations superimposed to the integer AB oscillation of
the persistent current and ring total energy.

\acknowledgments
This work has been performed under grants PB98-1247 
from DGESIC, Spain, and 2000SGR00024 from Generalitat de Catalunya.

\begin{figure}
\centering
\caption{
Addition energies $\Delta_2(N)$ (meV) at $B=0$
as a function of the number of electrons in the ring.
}
\label{fig1}
\end{figure}
\begin{figure}
\centering
\caption{
Electron density (solid line) for the $N=10$ ring at $B=0$.
The dimensionless horizontal scale may be transformed into a
conventional one recalling that $q=0.05$ nm$^{-1}$. The value
of $J_L(qr)$ for $L=0$ to 3 is also shown (right scale).
}
\label{fig2}
\end{figure}
\begin{figure}
\centering
\caption{
 Some results for the $N=10$ ring as a function of $B$. Top panel:
dipole CDE's (filled symbols, solid lines) and  SDE's (empty symbols,
dashed lines) in the long wave-length limit. The circles represent outer
edge modes, and the squares inner edge modes. Diamonds: bulk modes.
Middle panel: sp level energies. The $l$ value is indicated, and for
each of them, the lower energy level corresponds to the spin-up
orbital. The dashed line
marks the boundary between occupied and empty states. Bottom
panel: total energy (solid line, left scale), and persistent current
(dotted line, right scale in effective atomic units).
The values of the total spin and orbital angular momentum of the
different gs phases are also indicated.
}
\label{fig3}
\end{figure}
\begin{figure}
\centering
\caption{
 Referring to the $N=9$ ring, we display as a function of $B$:
Top panel, the sp level energies. The thick dashed line
marks the boundary between occupied and empty states.
Spin-up sp orbitals are denoted by solid lines, and spin-down sp
orbitals by dashed lines.
Bottom panel: total energy (solid line, left scale) and persistent
current (dotted line, right scale in effective atomic units).
The values of the total spin and orbital angular momentum of the
different gs phases are also indicated.
}
\label{fig4}
\end{figure}
\begin{figure}
\centering
\caption{
Electro-chemical potentials $\mu(N)=E(N)-E(N-1)$
as a function of $B$ . Different symbols are used to distinguish
gs's with different $S_z$ values. There can be
seen paramagnetic regions with $S_z=0$ (open circles), and a full
spin-polarized  region with $S_z=\slantfrac{N}{2}$
(black triangles). Other gs's with a  spin value
 $S_z=\slantfrac{1}{2}$ (grey  circles) and $S_z=1$
(black circles) may be identified. The region
of steadily increasing polarization has been indicated
by open triangles. The lines have been drawn to guide the eye.
}
\label{fig5}
\end{figure}
\begin{figure}
\centering
\caption{
Charge (solid lines) and spin density (dashed lines) strengths
in arbitrary units as a function of $\omega$ for
the $N=10$ ring at $B=0$ and several $q$ values. The
multipolarity of the peaks is indicated.
}
\label{fig6}
\end{figure}
\begin{figure}
\centering
\caption{
Charge density strength in arbitrary units
for $N=10$, $q=0.05$ nm$^{-1}$ and different $B$ values.
The multipolarity and polarization of the main peaks is indicated.
}
\label{fig7}
\end{figure}
\begin{figure}
\centering
\caption{
Same as Fig. \ref{fig7} for the spin density strength.
}
\label{fig8}
\end{figure}
\begin{figure}
\caption{$B$ dispersion of the more intense CDE's of the $N=10$ ring
for $q=0.05$ nm$^{-1}$. The polarization of the peaks is indicated, and
the lines have been drawn to guide the eye.
}
\label{fig9}
\end{figure}
\begin{figure}
\caption[]{Same as Fig. \ref{fig9} for the more intense SDE's.
}
\label{fig10}
\end{figure}
\begin{figure}
\centering
\caption{
Top panel: $\mu(10)$ (meV) around $B\sim 3$ T. The solid symbols represent
the ferromagnetic phase, and the open symbols the paramagnetic phase.
Bottom panel:
Charge (solid lines) and spin density (dashed lines) dipole strengths
in arbitrary units as a function of $\omega$ (meV) for the $N=10$
ring at $q=2.3 \times 10^{-3}$ nm$^{-1}$ around $B\sim 3$ T.
}
\label{fig11}
\end{figure}
\begin{figure}
\centering
\caption{
Grey-scale plot of CDE's (left panel) and SDE's (right panel) of
the $N=10$ ring for $q=0.05$ nm$^{-1}$ in the $B\sim 5-6$ T region.
The lines have been drawn to guide the eye. The dashed line along
one of the $1^-$ SDE's denotes an excitation branch
arising from the coupling
between charge and spin density responses.
}
\label{fig12}
\end{figure}

\begin{references}

\bibitem{Gar97} J.M. Garcia et al., Appl. Phys. Lett. {\bf 71}, 2014 (1997).

\bibitem{Lor98} A. Lorke and R. J. Luyken,
 Physica {\bf B} 256, 424 (1998).

\bibitem{Lor00}
A. Lorke. R. J. Luyken, A. O. Govorov, J. P. Kotthaus,
J. M. Garcia, and P. M. Petroff, Phys. Rev. Lett. {\bf 84}, 2223 (2000).

\bibitem{Dah93} C. Dahl, J. P. Kotthaus, H. Nickel, and W. Schlapp,
 Phys. Rev. B {\bf 48}, 15480 (1993).

\bibitem{Pro92} C. R. Proetto, Phys. Rev. B {\bf 46}, 16174 (1992).

\bibitem{Zar96} E. Zaremba, Phys. Rev. B {\bf 53}, R10512 (1996).

\bibitem{Cha94} T. Chakraborty and P. Pietil\"ainen, Phys. Rev. B
{\bf 50}, 8460 (1994).

\bibitem{Gud94} V. Gudmundsson and \'A. Loftsd\'ottir, Phys. Rev. B
{\bf 50}, 17433 (1994).

\bibitem{Wen96} L. Wendler, V. M. Fomin, A. V. Chaplik, and
A. O. Govorov, Phys. Rev. B {\bf 54}, 4794 (1996).

\bibitem{Hal96} V. Halonen, P. Pietil\"ainen, and T. Chakraborty,
Europhys. Lett. {\bf 33}, 377 (1996).

\bibitem{Nie96} K. Niemel\"a, P. Pietil\"ainen, P. Hyv\"onen,
and T. Chakraborty, Europhys. Lett. {\bf 36}, 533 (1996).

\bibitem{Tan99} W.-C. Tan and J. C. Inkson,
 Phys. Rev. B {\bf 60}, 5626 (1999).

\bibitem{Vie00} S. Viefers, P. Singha Deo, S.M.
Reimann, M. Manninen and M. Koskinen, Phys. Rev. B {\bf 62}, 10668 (2000).

\bibitem{Mag99} I. Magnusdottir and V. Gudmundsson, Phys. Rev. B {\bf 60},
16591 (1999)

\bibitem{Emp99} A. Emperador, M. Barranco, E. Lipparini, M. Pi, and
Ll. Serra, Phys. Rev. B {\bf 59}, 15301 (1999).

\bibitem{Hu00} H. Hu, J-L. Zhu, and J-J Xiong,
Phys. Rev. B {\bf 62}, 16777 (2000).

\bibitem{Kos00} M. Koskinen, M. Manninen, B. Mottelson, 
and S. M. Reimann, Phys. Rev. B {\bf 63}, 205323 (2001).

\bibitem{Bor00} P. Borrmann and J. Harting,
Phys. Rev. Lett. {\bf 86}, 3120 (2001).

\bibitem{Pue01} A. Puente and Ll. Serra,
Phys. Rev. B {\bf 63}, 125334 (2001).

\bibitem{Mai93} D. Mailly, C. Chapelier, and A. Benoit,
Phys. Rev. Lett. {\bf 70}, 2020 (1993).

\bibitem{Emp00} A. Emperador, M. Pi, M. Barranco and A.
Lorke, Phys. Rev. B {\bf 62}, 4573 (2000).

\bibitem{McE91}
P. L. McEuen, E. B. Foxman, U. Meirav, M. A. Kastner,
Y. Meir, N. S. Wingreen, and S. J. Wind, Phys. Rev. Lett.
{\bf 66}, 1926 (1991).

\bibitem{Ash92}
R. C. Ashoori, H. L. Stormer, J. S. Weiner,
L. N. Pfeiffer, and K. W. West, Phys. Rev. Lett. {\bf 68},
3088 (1992); {\bf 71}, 613 (1993).

\bibitem{Tar96} S. Tarucha, D. G. Austing, T. Honda, R. J. van
der Hage, and L. P. Kouvenhoven, Phys. Rev. Lett. {\bf 77},
3613 (1996).

\bibitem{Loc96} D. J. Lockwood, P. Hawrylak, P. D. Wang, C. M. Sotomayor
Torres, A. Pinczuk, and B. S. Dennis,
Phys. Rev.  Lett. {\bf 77}, 354 (1996).

\bibitem{Sch96} C. Sch\"uller, G. Biese, K. Keller,
C. Steinebach, D. Heitmann, P. Grambow, and K. Eberl,  Phys. Rev.
B {\bf 54}, R17\ 304 (1996).

\bibitem{Sch98} C. Sch\"uller, K. Keller, G. Biese, E. Ulrichs, L. Rolf,
C. Steinebach, D. Heitmann, and K. Eberl,  Phys. Rev. Lett. {\bf 80},
2673 (1998).

\bibitem{Fer94} M. Ferconi and G. Vignale, Phys. Rev. B
{\bf 50}, 14722 (1994).

\bibitem{Pi98} M. Pi, M. Barranco, A. Emperador, E. Lipparini, and Ll.
Serra, Phys. Rev. B{\bf 57}, 14783 (1998).

\bibitem{Ser99} Ll. Serra, M. Barranco, A. Emperador, M. Pi, and
E. Lipparini, Phys. Rev. B {\bf 59}, 15290 (1999).

\bibitem{Bar00}
M. Barranco, L. Colletti, E. Lipparini, A. Emperador, M. Pi, and
Ll. Serra, Rev. B {\bf 61}, 8289 (2000).

\bibitem{Fri96} M. Fricke, A. Lorke, J. P. Kotthaus, G. Medeiros-Ribeiro,
and P. M. Petroff, Europhys. Lett. {\bf 36}, 197 (1996).

\bibitem{Lor96} A. Lorke, M. Fricke, B. T. Miller, M. Haslinger,
J. P. Kotthaus, G. Medeiros-Ribeiro, and P. M. Petroff,
Proceedings of the 23rd Int. Symposium on Compound
Semiconductors, St. Petersburg (1996); Inst. Phys. Conf. Ser.
{\bf 155}, Chap. 11, pp. 803-808.

\bibitem{Mil97} B. T. Miller, W. Hansen, S. Manus, R. J. Luyken,
 A. Lorke, and J. P. Kotthaus,
Phys. Rev. B {\bf 56}, 6764 (1997).

\bibitem{Ste98}
O. Steffens, M. Suhrke, and  U. R\"ossler,
Europhys. Lett {\bf 44}, 222 (1998).

\bibitem{Tar00} S. Tarucha, D. G. Austing, Y. Tokura, W. G.  van
den Wiel, and L. P. Kouvenhoven, Phys. Rev. Lett. {\bf 84},
2485 (2000).

\bibitem{Vig87}
G. Vignale and M. Rasolt, Phys. Rev. Lett {\bf 59}, 2360 (1987);
Phys. Rev.  B {\bf 37}, 10 685 (1988).

\bibitem{Mos00} M. Moskalets, Physica B {\bf 291}, 350
(2000).

\bibitem{Pin89}  A. Pinczuk and G. Abstreiter, {\em Light
Scattering
in Solids V}, edited by M. Cardona and G. Guntherodt, Topics
in Applied Physics Vol. 66 (Springer-Verlag, Berlin, 1989) p. 153.

\bibitem{Stei99}
C. Steinebach, C. Sch\"uller, and D. Heitmann,
Phys. Rev.  B {\bf 59}, 10240 (1999).

\bibitem{Blu70} F. A. Blum, Phys. Rev. B {\bf 1}, 1125 (1970).

\bibitem{Ste99} O. Steffens and M. Suhrke,
Phys. Rev. Lett. {\bf 82}, 3891 (1999).

\bibitem{Emp01} A. Emperador, M. Pi, M. Barranco, E. Lipparini,
and Ll. Serra, Jpn. J. Appl. Phys. {\bf 40}, 518 (2001).


\end{references}
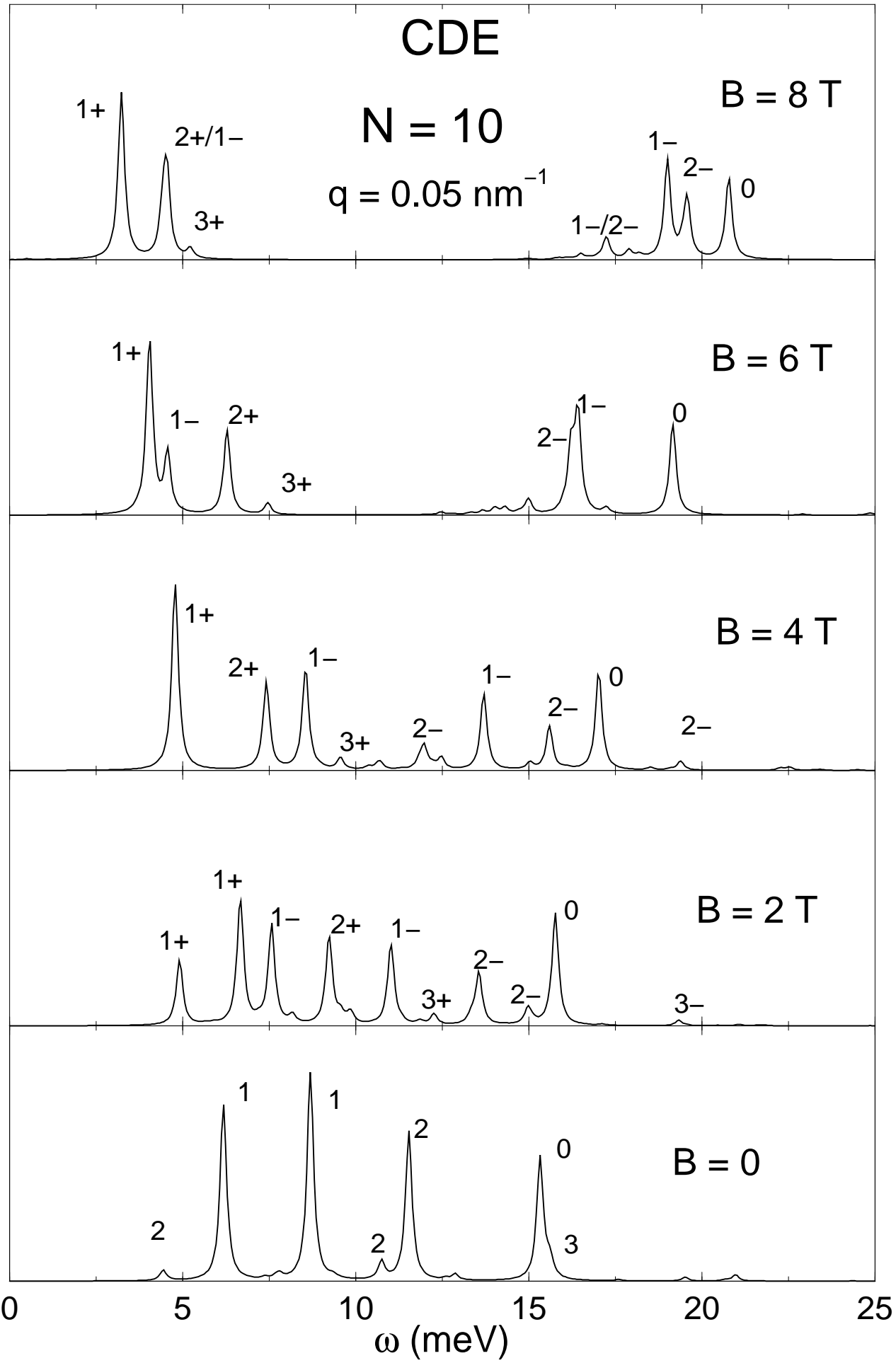
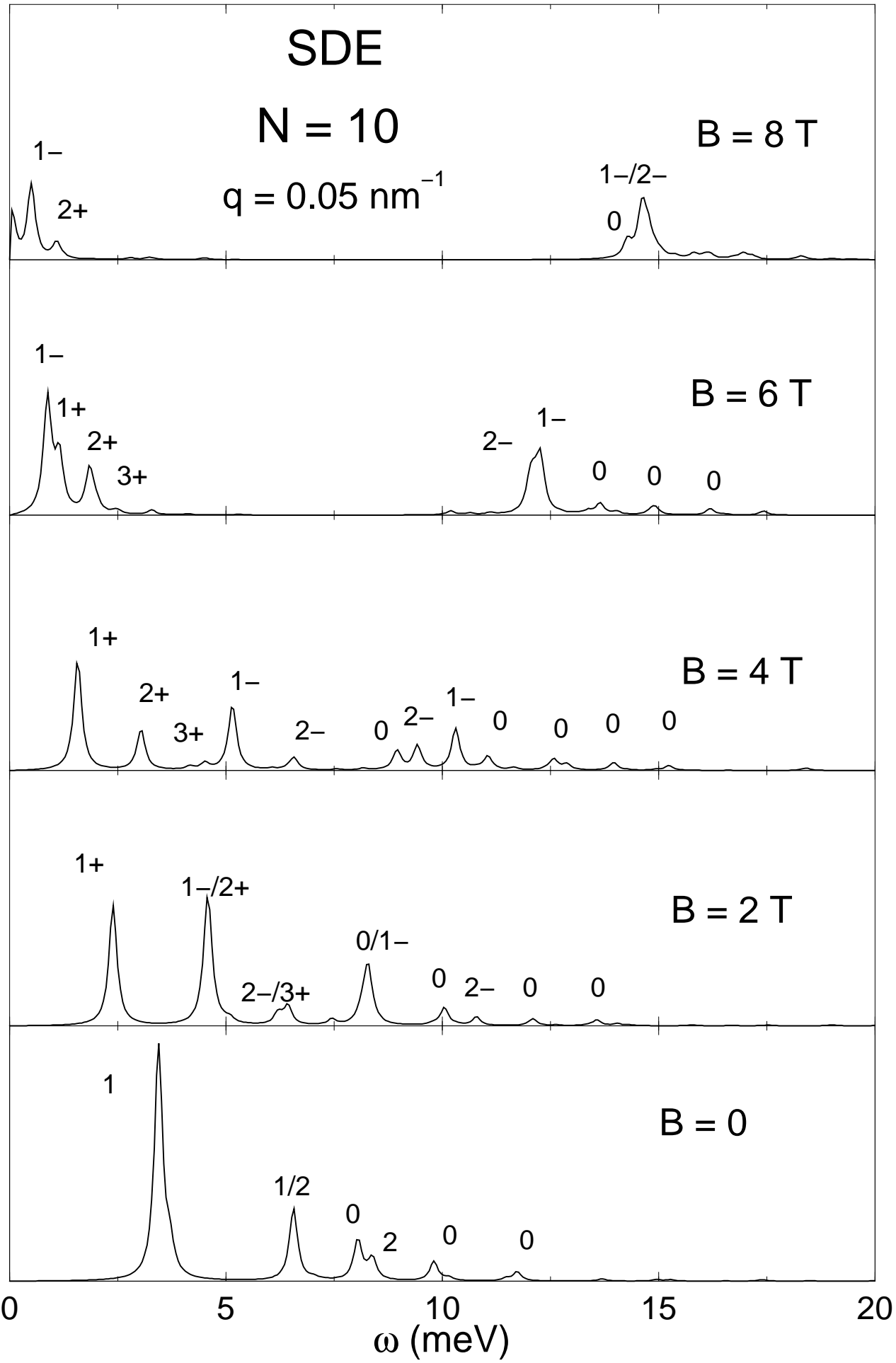
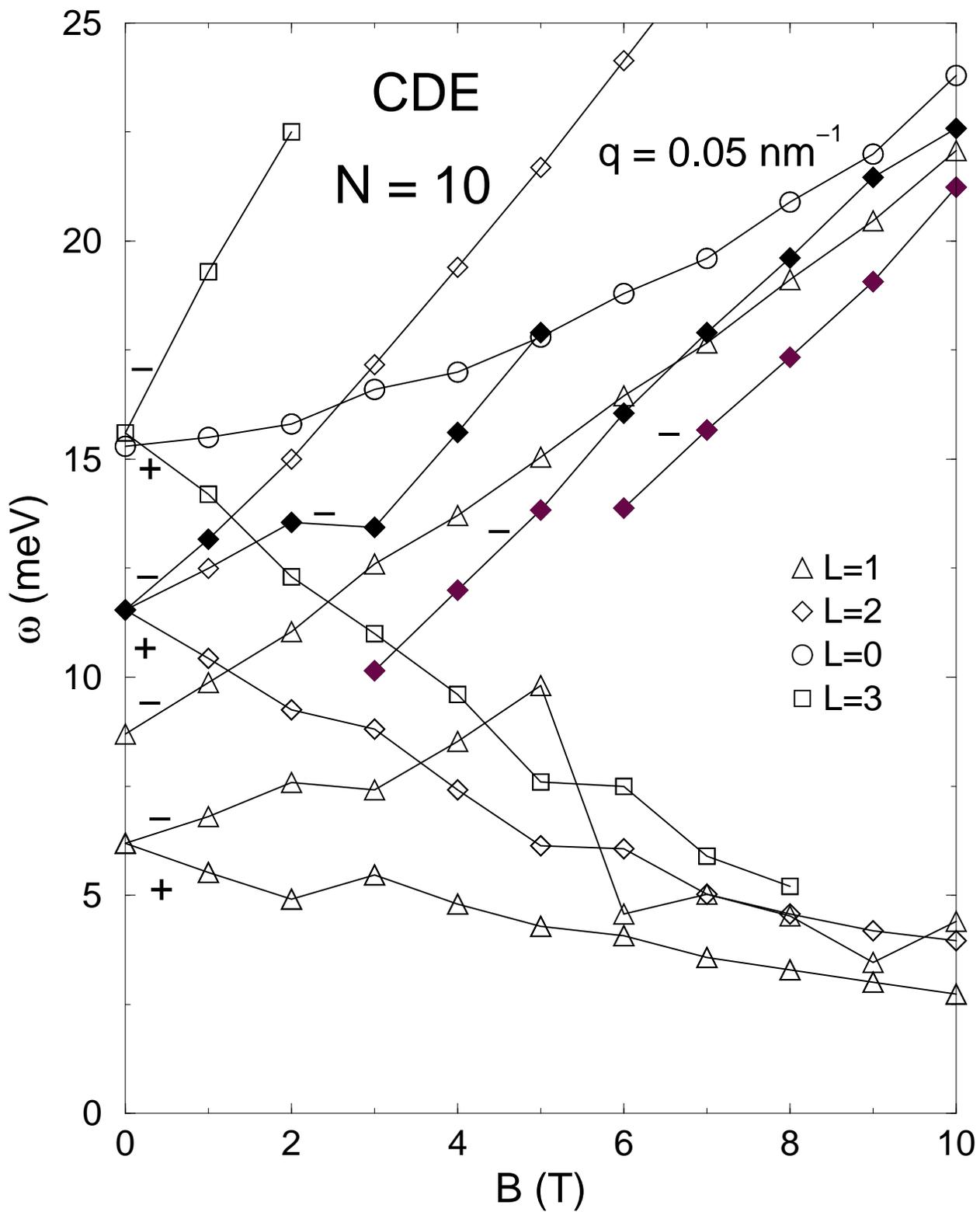
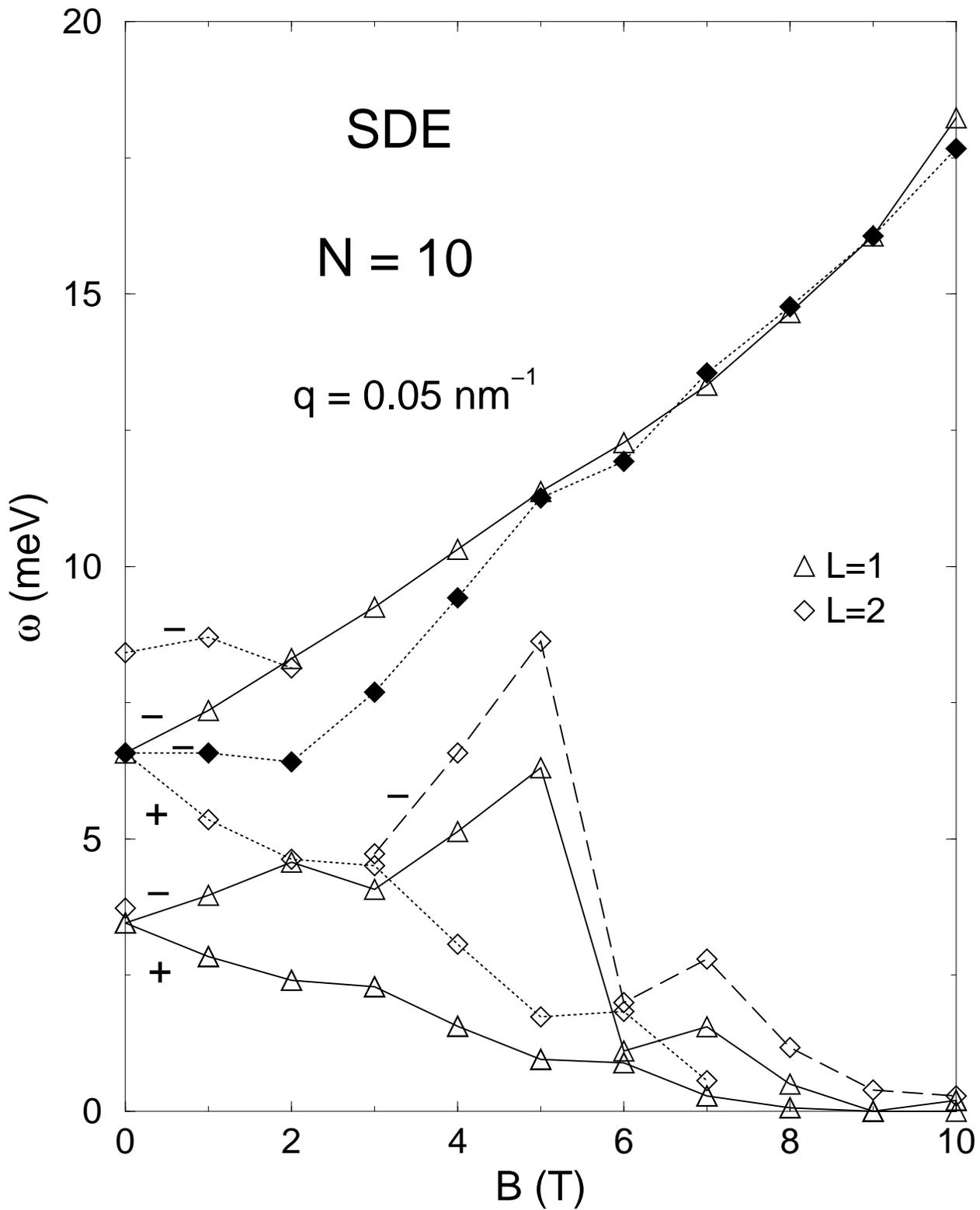
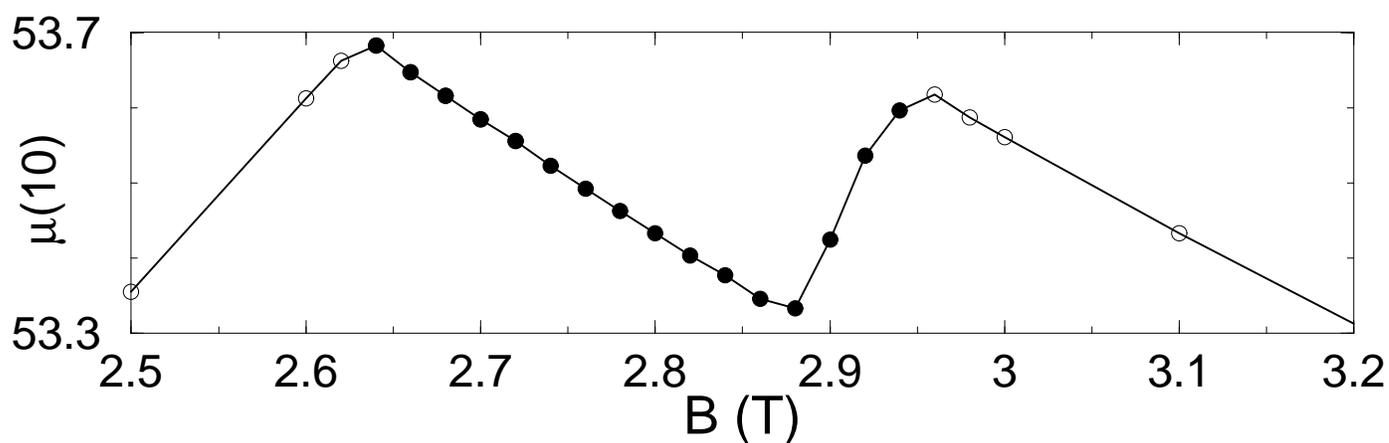
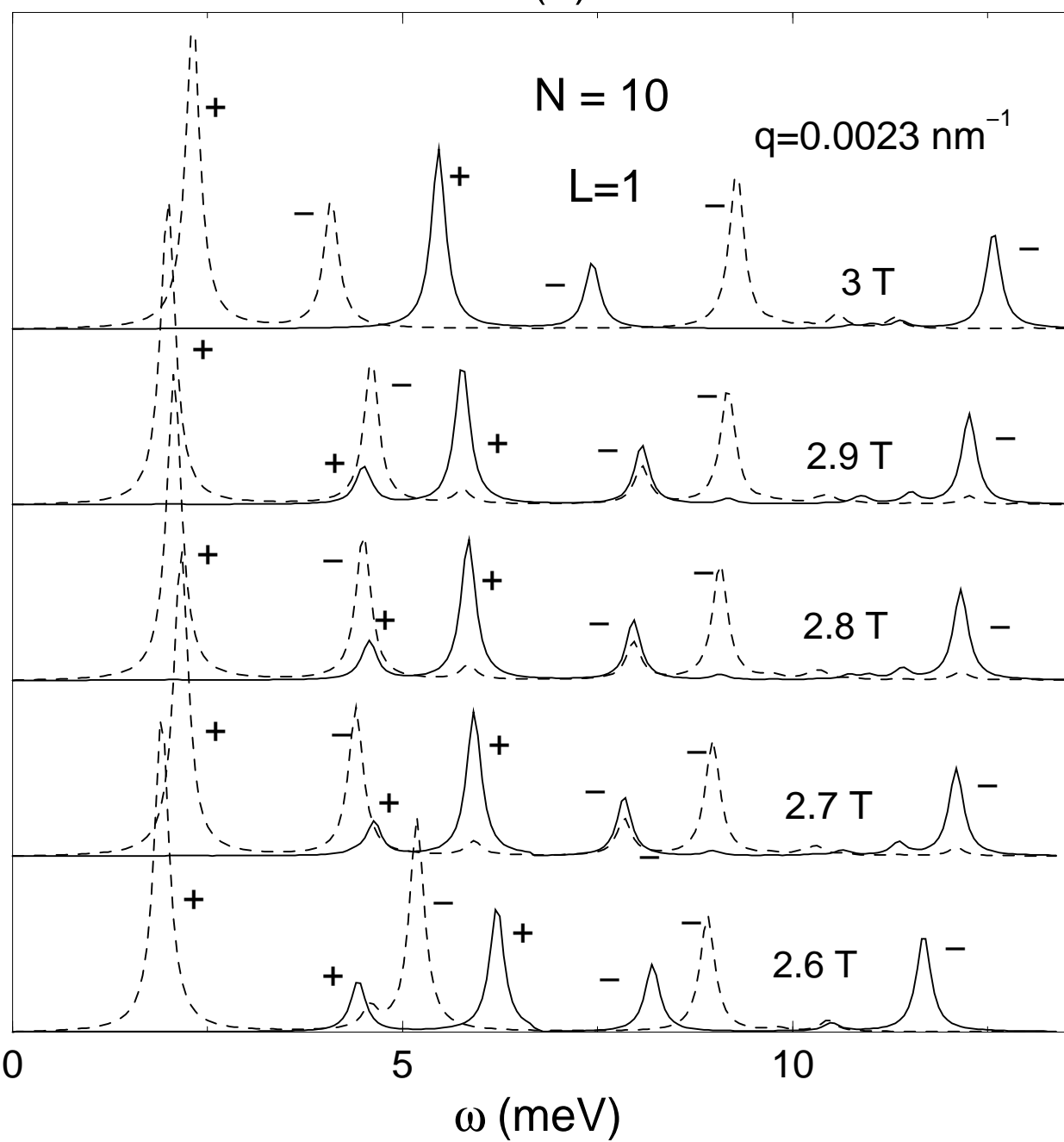
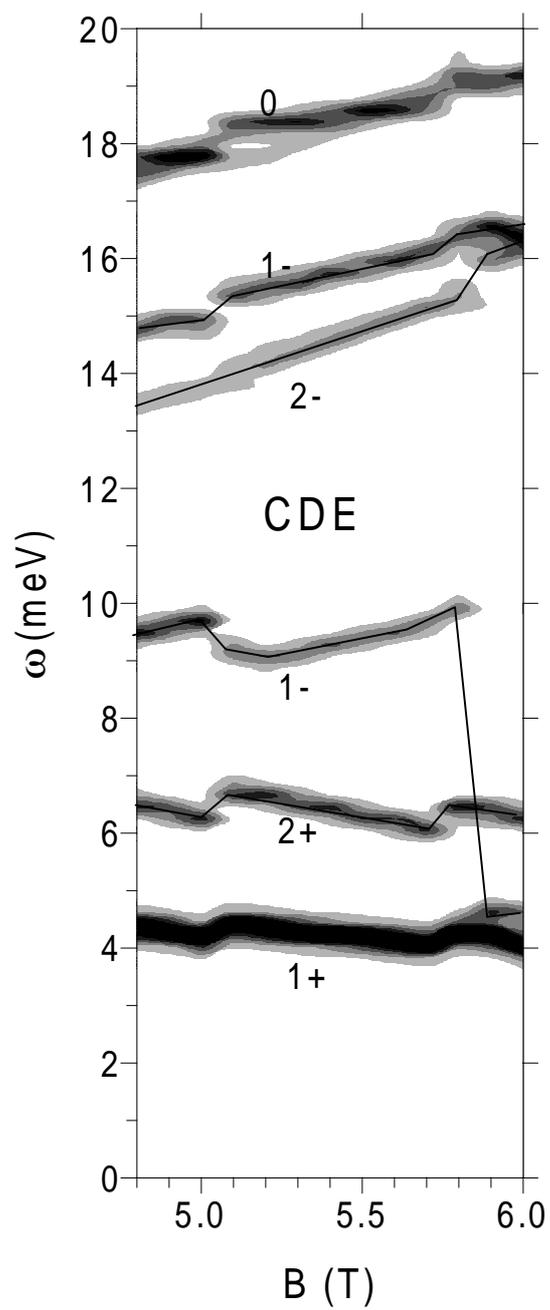
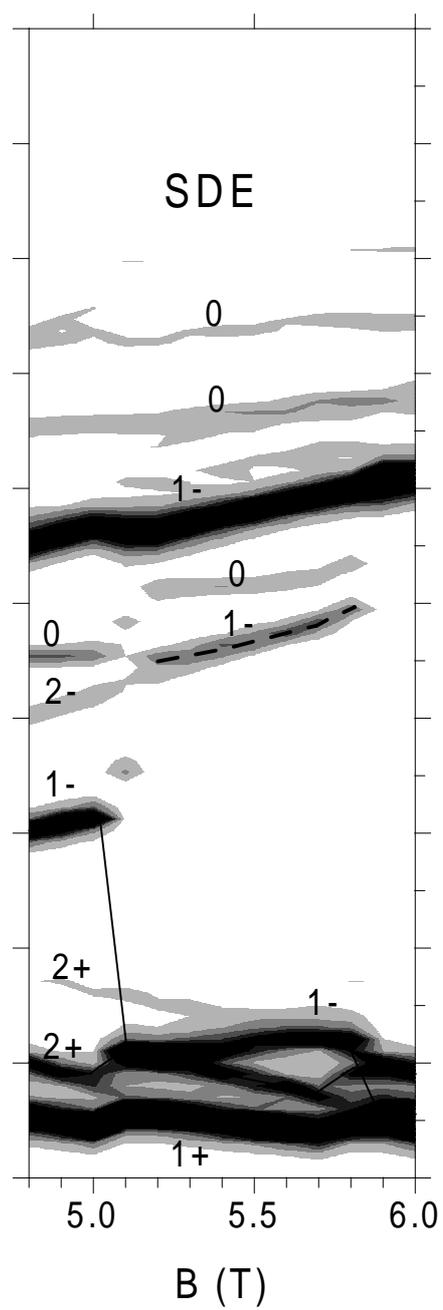
\end{document}